
\magnification \magstep1
\raggedbottom
\openup 2\jot
\voffset6truemm
\pageno 1
\headline={\ifnum\pageno=1\hfill\else
\hfill{\it Inflationary solutions in quantum cosmology}
\hfill \fi}
\centerline {\bf INFLATIONARY SOLUTIONS IN}
\centerline {\bf QUANTUM COSMOLOGY}
\vskip 1cm
\centerline {Giampiero Esposito$^{1}$
and Giovanni Platania$^{2}$}
\vskip 1cm
\noindent
{\it ${ }^{1}$Department of Applied Mathematics and
Theoretical Physics, University of Cambridge, Silver
Street, Cambridge CB3 9EW, UK}
\vskip 0.3cm
\noindent
{\it ${ }^{2}$Department of Physics, Mostra d'Oltremare,
Padiglione 19, 80125 Naples, Italy}
\vskip 1cm
\noindent
{\bf Abstract.}
We prove that in the Hartle-Hawking approach to quantum cosmology the
existence of an inflationary phase is a general property of minisuperspace
models given by a closed Friedmann-Robertson-Walker universe containing a
massless scalar field with a $\lambda\phi^{n}$ self-interaction. The evolution
in time of the cosmic scale factor and of the scalar field in the very early
universe is derived, together with the conditions to be satisfied in order
to solve the horizon and flatness problems.
\vskip 100cm
\leftline {\bf 1. Introduction}
\vskip 1cm
\noindent
The quantum state of the universe [1, 2] is described by a functional
$\psi$ of the 3-geometry and of the matter field configuration
on a compact spacelike 3-surface $S$. Any solution of the
Wheeler-DeWitt equation and of the momentum constraints is a
possible quantum state of the universe [1-3]. In the last few
years there has been great interest in the Hartle-Hawking
proposal for the boundary conditions of the universe
[1, 2, 4, 5], which seems to predict a universe whose
large-scale features are in agreement with observations
[1, 6-9]. According to Hartle and Hawking it seems that $\psi$
is to be taken over all compact Euclidean 4-metrics which
induce the given 3-metric on their boundary $S$, and over
all field configurations which match the given value on
$S$ and are regular on the compact 4-manifolds having $S$
as their only boundary.

In order to solve the Wheeler-DeWitt equation, most authors
have considered the minisuperspace approximation [10]. In the
by now well known Hawking massive scalar field model
[1, 2, 11, 12], the universe begins its inflationary
expansion from a non-singular state. At the end of the
inflation, it undergoes a matter- or radiation-dominated
expansion, reaches a maximum radius and recollapses to a
singularity. However, in general it is not clear whether
the solutions to the classical field equations corresponding
to the Hartle-Hawking wavefunction describe a universe
which will recollapse to a singularity. In fact, Louko [13]
has pointed out that the method of providing initial
conditions for purely Lorentzian solutions by analytically
continuing purely Euclidean solutions in the matter-dominated
and recollapsing phases may be erroneous. His calculations
seem to suggest that the universe will not reach the
singularity in the future.

After this short exposition of some ideas and problems
of quantum cosmology, let us now turn our attention to the
problem we try to solve in this paper. We study
a minisuperspace model given by a closed Friedmann-Robertson-Walker
(hereafter referred to as FRW)
universe containing a massless scalar field with a
$\lambda\phi^{n}$ self-interaction. A massless scalar field is considered for
two main reasons :
\item {(a)}
it is not clear whether a fundamental massive scalar field exists [7];
\item {(b)}
one is still unable to say which theory of particle physics gives the best
description of the very early universe [14].

In section 2 we write down the action integral
and the field equations of our
minisuperspace model, and we discuss in detail how the Hartle-Hawking
proposal enables one to impose initial conditions for the solutions to these
equations. We also endeavour to derive the oscillatory region for the wave
function in view of its behaviour in the limit of small 3-geometry
and by comparison with the massive scalar field model. In
addition we discuss how the
spacelike or timelike nature of the surfaces of constant potential of the
Wheeler-De Witt equation may be useful in deriving the behaviour
of the wavefunction.

In section 3 at first we focus our attention on the region in which the
wavefunction oscillates. The phase $S$ of the WKB approximation is
the one obtained
by analytical continuation of the Euclidean action for compact metrics and
regular matter fields. We are thus able to derive the evolution in time of
the cosmic scale factor and of the matter field using an approximate form
assumed by $S$. The Hartle-Hawking trajectories so derived are
exact solutions
to the Lorentzian field equations. They are distinguished with respect to the
general set of solutions in that they are inflationary and singularity-free
in the past.

In section 4 we derive the constraints to be satisfied in order
to solve the
horizon and flatness problems; in section 5 we summarize the results
obtained and we mention some related problems to be studied.
\vskip 1cm
\leftline {\bf 2. Minisuperspace model and boundary conditions}
\vskip 1cm
\noindent
Our model is based on the FRW metric which may be locally cast
in the form: $ds^{2}=
\sigma^{2}(-N^{2}dt^{2} + a^{2}(t)d\Omega^{2}_{3})$, where $\sigma =
\sqrt{2G \over 3\pi}$, $N$ is the lapse function [15],
$a(t)$ is the cosmic scale
factor and $d\Omega^{2}_{3}$ is the metric on a unit 3-sphere.

Let $\phi'(t) = {\phi(t) \over \pi\sigma\sqrt{2}}$
be the scalar field (in these
units $\phi$ is dimensionless) having a potential of the form $V(\phi') =
{\lambda'{\phi'}^{n} \over 2}$. The action integral containing
a surface term [5] becomes, in our case,
$$
I=-{1\over 2} \int Na^{3} \left[{{\dot a}^{2} \over a^{2}N^{2}}
- {1\over a^{2}}
-{\dot \phi^{2} \over N^{2}} + \lambda\phi^{n}\right]dt = \int L\ dt
\eqno (2.1)
$$
where $\lambda = \lambda'\sigma^{-n+4}(\pi\sqrt{2})^{-n+2}$.

Before commencing the calculations, it is worth emphasizing the main property
of minisuperspace models containing scalar fields. The minisuperspace model
given by a homogeneous isotropic universe without matter fields but with a
positive cosmological constant (De Sitter space) did show that the
Hartle-Hawking proposal leads to inflation if some mechanism is able to give
rise to an effective cosmological constant in the early universe [1, 11].
But our universe is not expanding exponentially.
This is why we require that the parameter
we call the cosmological constant has finally to vanish as the time
passes. The simplest model for generating a decaying cosmological constant is
the one containing a massive scalar field. In fact, when $\phi$ is initially
very large and almost constant, the $m^{2}\phi^{2}$ term in the action of the
massive model acts as an effective cosmological constant. This implies that,
when $\phi\cong$constant=$\phi_{0}$, all formulae for the massive model can
be obtained from the De Sitter model by putting $H=\sqrt{m^{2}\phi_{0}^{2}}=
m\phi_{0}$, where $H$ is the parameter of the De Sitter model related to the
cosmological constant $\Lambda_{c}$ by means of the relation [1]
$H^{2}={2G\Lambda_{c} \over 9\pi}$. If $a<H^{-1}$, it is known that there
are two solutions of the Euclidean field equations, and the dominant
contribution to the semiclassical approximation of the wave function comes
from the one having action [1]
$I_{E}=-{1 \over 3H^{2}}\Bigr[1-(1-H^{2}a^{2})^{3\over 2}\Bigr]$,
where $I_{E}$ is the action of the smaller part of a 4-sphere of radius
${1\over H}$ bounded by a 3-sphere of radius $a$. In our model, if the
term $\lambda\phi^{n}$ in (2.1) has to act as an effective cosmological
constant when $\phi$ is large and constant, the parameter $H$ becomes $H=
\sqrt{\lambda\phi_{0}^{n}}=\sqrt{\lambda}\phi_{0}^{n \over 2}$. From now on,
$H$ will always denote $\sqrt{\lambda}\phi_{0}^{n \over 2}$ in our equations.

We can now turn again our attention to (2.1). Putting $N=1$, the Lorentzian
field equations are
$$
a \ddot a+{{\dot a}^{2} \over 2}+{1 \over 2}+{3 \over 2}a^{2}
{\dot \phi}^{2}-{3 \over 2}\lambda a^{2}\phi ^{n} = 0
\eqno (2.2)
$$
$$ \ddot \phi + 3{\dot a \over a}\dot \phi + {n \over 2}\lambda
\phi ^{n-1} = 0 .
\eqno (2.3)
$$
We now propose to derive the initial conditions for the solutions to (2.2)
and (2.3) implied by the Hartle-Hawking proposal. To this purpose let us
recall that the Hartle-Hawking ground state is a Euclidean functional
integral taken over compact 4-metrics and regular matter fields. In the
Euclidean regime one thus has [1]
$$
a(\tau=0)=0\; \; \; \; \; \; \; \; \; \; \; \; \; \; \;
\dot a(\tau=0)=1
\eqno (2.4)
$$
$$
\phi(\tau=0) =\phi_{0}\; \; \; \; \; \; \; \; \; \; \; \; \; \; \;
\dot \phi(\tau=0)=0 .
\eqno (2.5)
$$
In fact, having taken the path integral over compact 4-metrics, $a(\tau)$
has to vanish at some value of $\tau$ we can choose to be zero [1].
This implies that at large values of $\phi$ one has $a(\tau) = {1 \over
H}\sin(H\tau)$, and this relation is in agreement
with (2.4). In addition, in the Euclidean regime (2.3) can be
written in a
similar form, provided we take the derivatives with respect to $\tau$ and we
change the sign in front of ${n\over 2}\lambda\phi^{n-1}$.
Therefore, multiplying both sides of (2.3) by $a$, and
letting $a$ go to zero, one has : $ a\ddot \phi \rightarrow 0$, ${n\lambda
\over 2}a\phi ^{n-1} \rightarrow 0$, in view of the regularity of $\phi$.
Therefore one has $a(3{\dot a \over a}\dot \phi)=0$ if and only if $\dot
\phi=0$, in view of (2.4).

In quantum cosmology one is mainly interested in the oscillatory regions for
the wavefunction, because such regions correspond to classically allowed
regions [1].
One can then make the WKB ansatz: $\psi= \sum Re[C_{n}\exp
(iS_{n})]$ (in what follows we shall consider a single component of the wave
packet). The phase $S$ is chosen to satisfy the Hamilton-Jacobi equation for
general relativity, and the Hartle-Hawking proposal picks out the
Hamilton-Jacobi function which is the analytic continuation of the Euclidean
action. This implies that the Lorentzian Hartle-Hawking trajectories are
obtained by analytic continuation of the compact Euclidean paths with regular
matter fields. The application of this method yields,
putting $\tau = {\pi \over 2H}+it$ [12]:
$$
a(t)={1 \over H}\cosh (Ht)
\eqno (2.6)
$$
$$
\phi(t) = \phi_{0}
\eqno (2.7)
$$
at very small values of the time. Thus the minimum value of $a$ in the
Lorentzian regime
is equal to the maximum value of $a$ in the Euclidean regime, and
the initial conditions for $a$ differ greatly from (2.4) because one has
$a(t=0)={1 \over H}$, $\dot a(t=0)=0$.
Therefore at large initial values of $\phi$,
the initial conditions for the solutions
to (2.2) and (2.3) are (see also
(3.3) and the subsequent discussion):
$$
\phi(t=0)=\phi_{0}\; \; \; \; \; \; \; \; \; \; \; \; \; \; \;
a(t=0)={1 \over \sqrt{\lambda}\phi_{0}^{n \over 2}}=a_{0}
\eqno (2.8)
$$
$$
\dot \phi(t=0)=0\; \; \; \; \; \; \; \; \; \; \; \; \; \; \;
\dot a(t=0)=0 .
\eqno (2.9)
$$
Moreover, putting [8]
$\alpha=\log (a)$, $p_{a}^{2}=-{1 \over a}{\partial({{a}
{\partial \over \partial a}})
\over \partial a}$, the Wheeler-De Witt equation for our model becomes
$$
{1 \over 2}N e^{3\alpha}
\left[{\partial ^{2} \over \partial \alpha^{2}}-
{\partial ^{2} \over \partial \phi^{2}}+
(\lambda\phi^{n}e^{6\alpha}  -e^{4\alpha})\right]
\psi(\alpha,\phi)=0 .
\eqno (2.10)
$$

Now, the action $I$ of a compact solution of the field equations
vanishes in the limit of small 3-geometries and one often argues
that the prefactor $A$ of the semiclassical approximation of
the wavefunction, $\psi \sim A \; \exp(-I)$ (this is the formula
for the solution of the Wheeler-DeWitt equation to leading order),
may be taken to be a constant, so as to normalize $\psi$ to $1$
in the limit of small 3-geometries [8, 16]. This would be a
boundary condition for (2.10) implied by the Hartle-Hawking
proposal, and in numerical calculations it has been used so as
to derive the oscillatory or exponential regions
of the wavefunction [16]. Thus, from the knowledge of $\psi$
at small 3-geometries, one derives its behaviour at larger
3-geometries [16].

Strictly speaking, this is not completely true, because the
prefactor of the semiclassical approximation is given by
the path integral of $\exp(-I_{2})$, where $I_{2}$ is the
part of the action quadratic in the fluctuations about
classical solutions. This calculation is not trivial and
one can show, for example, that in the case of pure gravity
the prefactor diverges in the limit of small 3-geometries
in the Euclidean regime [17]. Thus, in general, the
normalization of $\psi$ to $1$ may not be sufficiently
accurate. However, in our opinion one can still take, to
a first approximation, the point of view of Hawking and
Wu [16]. They point out that a suitable choice (also
made in equation (2.10) in this paper) for the factor
ordering in the Wheeler-DeWitt equation is consistent with
a constant behaviour of the prefactor and of the
wavefunction in the limit of small $a$, as can also be
verified in our case. In fact
$\Bigr({\partial^{2}\psi \over \partial \alpha^{2}}
-{\partial^{2}\psi \over \partial \phi^{2}} \Bigr)$ vanishes
if $\psi$ is a constant and the potential
$V(\alpha,\phi)=e^{6\alpha}\lambda \phi^{n}-e^{4\alpha}
=a^{6}\lambda \phi^{n}-a^{4}$, for a given value of $\phi$,
vanishes when $a \rightarrow 0$. The numerical integration
of the Wheeler-DeWitt equation with this condition shows
that in the massive model $\psi$ starts to oscillate when
$\mid \phi \mid$ is greater than one and $V(\alpha,\phi)$
is positive. In our case, it is clear that, even if one puts
$\lambda^{{1\over 2}}={\tilde m}$,
$\phi^{{n\over 2}}=\Phi$, one cannot cast (2.10) in a form
which is formally identical to the massive case, because
${\partial^{2}\psi \over \partial \phi^{2}}$
is very different from
${\partial^{2}\psi \over \partial \Phi^{2}}$.
However, in the case of a $\lambda\phi^{n}$ theory, for a
given value of $\phi$ one has an even larger range of values of $a$ for which
the potential V($a$,$\phi$) in the Wheeler-De Witt equation is positive,
if $\lambda$ is not too small. The
sign of this potential plays a crucial role in determining the oscillatory
behaviour of $\psi$ and this is a quite general result not depending on the
particular theory for the scalar field [18].
It is particularly useful to
discuss this matter in detail. To begin with, we point out that the
Wheeler-De Witt equation (2.10) is a hyperbolic equation in the
minisuperspace with metric $ds^{2}=e^{3\alpha}(-d\alpha^{2}+d\phi^{2})$,
and we consider the regions in which the surfaces of constant
$V=e^{6\alpha}\lambda\phi^{n}-e^{4\alpha}$ are spacelike with respect to
this metric. The idea is that in a local region we can perform a Lorentz
transformation [18] to new coordinates
($\widetilde{\alpha}$,$\widetilde{\phi}$) so that the constant-$V$
surfaces are
parallel to the $\widetilde{\phi}$ axis. This implies that (2.10) takes the
form $\Bigr[{\partial^{2} \over \partial\widetilde{\alpha}^{2}}-
{\partial^{2} \over
\partial\widetilde{\phi}^{2}}+V(\widetilde{\alpha})\Bigr]\psi=0$ and can be
solved by separation of variables, putting
$\psi=\psi_{1}(\widetilde{\alpha})\psi_{2}(\widetilde{\phi})$. One thus finds
the differential equation for $\psi_{1}(\widetilde{\alpha})$:
$\Bigr[{d^{2}\psi_{1} \over
d\widetilde{\alpha}^{2}}+\Lambda+V(\widetilde{\alpha})\Bigr]\psi_{1}
(\widetilde {\alpha})=0$, whose solution has oscillatory behaviour if
$\Lambda+V(\widetilde{\alpha})>0$, where $\Lambda$ is a separation constant.
It is also possible that the surfaces of constant $V$ are timelike. In such a
case they can be locally rotated so as to lie parallel to the
$\widetilde{\alpha}$ axis, which in turn implies that
$V=V(\widetilde{\phi})$. In such a case one finds the following differential
equation for $\psi_{2}(\widetilde{\phi})$: $\Bigr[{d^{2}\psi_{2} \over
d\widetilde{\phi}^{2}}
+\Lambda'-V(\widetilde{\phi})\Bigr]\psi_{2}(\widetilde{\phi})=0$ ,
whose solution has
oscillatory behaviour if $\Lambda'-V(\widetilde{\phi})>0$, where
$\Lambda'$ is another separation constant. Thus $V$ is expected to be negative
in this case, if the separation constant $\Lambda'$ does not assume too
large a positive value.

To sum up, if the $V={\rm constant}$
surfaces are spacelike in the metric of the
minisuperspace, $\psi$ is expected to
oscillate if V$>$0 and to behave exponentially if V$<$0, whereas the converse
should hold true if the $V={\rm constant}$
surfaces are timelike. In our model, the
nature of the surfaces of constant potential is determined by the quantity
$\Sigma=g^{ab}V_{,a}V_{,b}$, where
$g^{ab}=e^{3\alpha} \; {\rm diag}(-1,1)$. The sign
of this quantity shows the timelike or spacelike nature of the vector normal
to these surfaces. Computing $\Sigma$ when $V=k>0$, one finds
$$\eqalignno{
\Sigma&=e^{3\alpha}  \left[-(V_{,\alpha})^{2}+(V_{,\phi})^{2}\right]_{V=k>0}
\cr
&=-e^{3\alpha}  \left[\left(1-{n^{2} \over \phi^{2}}\right)
(k+e^{4\alpha})^{2}+2(k+e^{4\alpha})(5k+e^{4\alpha})+(5k+
e^{4\alpha})^{2}\right]<0
\cr}
$$
if $\phi$ or $e^{\alpha}$
are sufficiently large. This implies that the surfaces of constant
positive potential are spacelike for such values of $\phi$, which in turn
implies that the wavefunction has to oscillate. Honestly speaking,
in the massive scalar field model there are some regions in which this method
disagrees with numerical results for the wavefunction derived under the
assumption that $\psi=1$ when $a\rightarrow0$ [19].
In view of the fact that
both these methods have flaws, the question seems to us to be an open one.
We can thus show that the wavefunction of our model oscillates
when $V(\alpha,\phi)$ is positive in one of the following two ways:

(a) by comparison with the results derived in the massive model,
pointing out that also in our case one can take
$\psi \rightarrow 1$ as $a \rightarrow 0$, and that in our case
$V(\alpha,\phi)$ has an even stronger positivity property if
the self-coupling parameter $\lambda$ is not too small;

(b) by studying the nature of the surfaces
$V(\alpha,\phi)$=constant.

In the case of a $\lambda\phi^{4}$ theory, we are in agreement with the
result derived in a paper by Zhuk [20]. However in this paper the
normalizability of $\psi$ to 1 is not a boundary condition assumed from the
beginning, but a consequence of the mathematical derivation, in which one
requires that the solution of the Wheeler-De Witt equation has to vanish when
the cosmic scale factor $\rightarrow\infty$.

So far we have emphasized that, at almost
constant values of $\phi$, our model can be
obtained from the massive one replacing the positive quantity $m^{2}\phi^{2}$
by means of $\lambda\phi^{n}$. This is possible only when $\phi$ is positive,
or when $\phi$ is negative but $n$ is even. In view of
this result we shall not
take into account the case when $\phi$ is negative and $n$ is odd. Of course,
the initial value of $\mid\phi\mid$ is not completely arbitrary, because the
form of S involving $\phi$ must be a solution of the Hamilton-Jacobi
equation in the oscillatory region for the wavefunction. In our case $S$
satisfies the equation
$$
{\left({\partial S \over \partial \phi}\right)}^{2}-
{\left({\partial S \over \partial \alpha}\right)}^{2} =
e^{4\alpha}  -\lambda\phi^{n}e^{6\alpha}
\eqno (2.11)
$$
where the right-hand side of (2.11) is negative. A careful
examination of the
quantity $\Sigma$ shows that this is possible only if $\mid\phi\mid$ is
initially at least greater than a number of order one.
We have not solved numerically the
Wheeler-De Witt equation for our model, and this is an interesting problem to
be studied for further research. However, it is to be
remarked that the numerical techniques used so far have some limits. For
example, the leapfrog algorithm is a valid approximation only when the
potential $V$ is everywhere much smaller than the inverse of the square of the
grid step size [19].
\vskip 1cm
\leftline {\bf 3. Lorentzian Hartle-Hawking trajectories and
field equations}
\vskip 1cm
\noindent
In this section we shall perform at first our calculations for positive values
of $\phi$, and later on we shall also discuss briefly the case when $n$ is
even and $\phi$ is negative.

The Euclidean action for compact metrics and regular matter fields is given
in our model by
$I_{E}=-{1 \over 3\lambda\phi^{n}}\Bigr[1-(1- e^{2\alpha}\lambda
\phi^{n})^{3 \over 2}\Bigr]$. The geometrical interpretation of $I_{E}$
is that it
is the action of the smaller part of a 4-sphere of radius ${1 \over \sqrt
{\lambda}\phi^{n \over 2}}$, bounded by a 3-sphere of radius $
a=e^{\alpha}$; in so doing, we generalize what has been done in the case of
the massive model [1, 11],
as already emphasized in section 2. The solution of
(2.11) is, in the Hartle-Hawking approach, the analytical continuation of
$I_{E}$. At large values of $\phi$ it is given by
$$
S\cong -{1 \over 3\lambda\phi^{n}}
(e^{2\alpha}  \lambda\phi^{n}-1)^{3 \over 2} .
\eqno (3.1)
$$

In fact the insertion of
(3.1) into the left-hand side of (2.11) yields
$$ \eqalignno{
{\left({\partial S \over \partial \phi}\right)}^{2}-
{\left({\partial S \over \partial \alpha}\right)}^{2}&=
{n^{2}e^{4\alpha}   \over 4\phi^{2}}(e^{2\alpha}  \lambda\phi^{n}-1)+
{n^{2} \over 9\lambda^{2}\phi^{2n+2}}(e^{2\alpha}  \lambda\phi^{n}-1)^{3}\cr&-
{n^{2}e^{2\alpha}   \over 3 \lambda
\phi^{n+2}}(e^{2\alpha}  \lambda\phi^{n}-1)^{2}
+e^{4\alpha}  -\lambda\phi^{n}e^{6\alpha} &(3.2)\cr}
$$
and we can easily recognize that the right-hand side of (3.2) reduces to
$e^{4\alpha}-\lambda\phi^{n}e^{6\alpha}$ at large initial $\phi$ and
small $t$. The relation (3.1) shows again that $\phi$ cannot be
negative when $n$
is odd, because in such a case $S$ becomes complex, whereas by its very
definition $S$ is real. In addition we know
that the function $S$ in (2.11) and (3.1) defines
a first integral of the system:
$$
p_{a}={\partial S \over \partial a}=
{\partial L \over \partial \dot a}
$$
$$
p_{\phi}={\partial S \over \partial
\phi}= {\partial L \over \partial \dot \phi}
$$
where $L$ is the Lagrangian
defined in (2.1). This system therefore becomes, putting $N=1$
$$
\dot a = (a^{2}\lambda\phi^{n}-1)^{1 \over 2}
\eqno (3.3)
$$
$$
\dot \phi =
-{n \dot a \over 2a\phi}+{n \over 3\lambda\phi^{n+1}}
{\left({\dot a \over a}\right)}^{3} .
\eqno (3.4)
$$

It is to be remarked that this system is in full agreement with the initial
conditions (2.8) and (2.9) in the Lorentzian regime derived from the
Hartle-Hawking proposal. In fact, for example, $\dot a(t=0)=0$, when
substituted into (3.3), yields $a_{0}={1 \over \sqrt{\lambda}\phi_{0}^{n
\over 2}}$.
As the time passes, the dominant contribution to $a(t)$ is given by
the increasing exponential, so that $a(t) \cong {\exp(Ht) \over 2H}$. Under
such assumptions, (3.1) takes the approximate form:
$$
S \cong -{e^{3\alpha}  \sqrt{\lambda}\phi^{n \over 2} \over 3} .
\eqno (3.5)
$$

The relation (3.1) enables us to also compute the first
correction to (3.5)
involving negative powers of $\phi$. In fact, from (3.1) one has that
$$
\eqalignno{
S&\cong
-{1 \over 3\lambda\phi^{n}}
\left[\sqrt{e^{2\alpha}  \lambda\phi^{n}\left(1 - {1 \over
\lambda e^{2\alpha} \phi^{n}}\right)}\right]^{3}\cr
&\cong-{1\over 3\lambda \phi^{n}}(e^{\alpha}  \sqrt{\lambda}\phi^{n\over
2})^{3}\left(1-{1 \over 2\lambda e^{2\alpha} \phi^{n}}\right)^{3}\cr
&\cong
-{\sqrt{\lambda} \over 3}e^{3\alpha}  \phi^{n \over 2}\left(1-
{3 \over 2\lambda e^{2\alpha} \phi^{n}}\right) . &(3.6)\cr}
$$

In the derivation of (3.6), we have used the following approximate formulae
which are valid when $0<\mid x \mid\ll 1$:
$$
\sqrt{1+x}\cong1+{x \over 2};\; \; \; \; (1+x)^{n}\cong 1+nx
\eqno (3.7)
$$
where in our case $x=-{1 \over \lambda\exp(2\alpha)\phi^{n}}$. The derivation
of (3.6) shows that (3.5) holds true when $\phi$ is much greater
than one or when $e^{\alpha}$ is very large.

Let $[0,t_{a}[$ be the time interval during which the Lorentzian
Hartle-Hawking trajectories (hereafter referred to as LHH, see [13])
evolve according to (2.6), (2.7), (3.3) and (3.4).
We are now interested in deriving solutions of the system
$$
p_{a}=
{\partial S \over \partial a}={\partial L \over \partial \dot a}
$$
$$
p_{\phi}=
{\partial S \over \partial \phi}={\partial  L \over \partial \dot \phi}
$$
in the interval $[t_{a},t_{b}]$ during which $S$ can be approximated as in
(3.5). Inserting (3.5) into this system,
and putting $N=1$, one finds that
$$
\dot \phi = -{n\sqrt{\lambda} \over 6}\phi^{{n \over 2}-1}
\eqno (3.8)
$$
$$
\dot \alpha = \sqrt{\lambda}\phi^{n \over 2} .
\eqno (3.9)
$$

By integrating (3.8)
and putting $\phi_{a}=\phi(t_{a})$, where $\phi_{a}=\phi_
{0}$, we get
$$
\phi(t)=\left[\phi_{a}^{{(4-n) \over 2}}-\left(2-{n \over 2}\right)
{n \sqrt{\lambda} \over 6}
(t-t_{a})\right]^{2 \over (4-n)}
\eqno (3.10)
$$
for every $n \not = 4$. This implies
$$
a(t)=a_{I}\exp{\left\{\sqrt{\lambda}\int _{t_{a}}^{t}
{\Bigr[\phi_{a}^{{(4-n)\over 2}}
-\left(2-{n \over 2}\right){n\sqrt{\lambda}\over 6}(t'-t_{a})\Bigr]}
^{{n \over (4-n)}}dt'\right\}}
\eqno (3.11)
$$
for every $n \not = 4$, where $a_{I}={1 \over H}\cosh(Ht_{a})\cong {1 \over
2H}\exp(Ht_{a})$. In order to study the case $n>4$, we define $k=\phi_{a}^
{{(4-n) \over 2}}$, $b=({n \over 2}-2){n\sqrt{\lambda} \over 6}$, $z={n
\over n-4}$. Therefore (3.11) becomes
$$
a(t)=a_{I}\exp{\left\{{\sqrt{\lambda}\over b(-z+1)}\Bigr[(k+b(t-t_{a}))
^{(-z+1)} - k^{(-z+1)}\Bigr]\right\}} .
\eqno (3.12)
$$

The second term in the curly brackets of (3.12) is
$$
{\sqrt{\lambda} \over b(z-1)}k^{(-z+1)}={3\phi_{a}^{2} \over n} .
\eqno (3.13)
$$
Moreover, at the end of the era during which (3.12) holds, and putting
$\phi_{b}=\phi(t_{b})$, one has from (3.10) that
$$
t_{b}-t_{a}={12 \over n(n-4)\sqrt{\lambda}}\phi_{a}^{{(4-n) \over 2}}
\left[{\left({\phi_{b}\over \phi_{a}}\right)}^{{(4-n) \over 2}}-1\right]
\cong {12 \over n(n-4)\sqrt{\lambda}}\phi_{b}^{{(4-n) \over 2}}
\eqno (3.14)
$$
if $\phi_{b}\ll \phi_{a}$. Thus the first term in the curly brackets of
(3.12) becomes
$$
{\sqrt{\lambda} \over b(-z+1)}(k+b(t_{b}-t_{a}))^{(-z+1)}\cong
-{3\phi_{b}^{2} \over n} .
\eqno (3.15)
$$

{}From (3.13) and (3.15) we get
$$
a(t_{b})=a_{I}\exp{\left[{3\phi_{a}^{2} \over n}
\left(1-{\Bigr({\phi_{b}\over \phi_{a}}\Bigr)}^{2}\right)
\right]}\cong a_{I}\exp{\left({3\phi_{a}^{2}
\over n}\right)} .
\eqno (3.16)
$$

If $n=4$, (3.10) and (3.11) are no longer valid; in such a case, the
integration of (3.8) yields
$$
\phi = \phi_{a}\exp{\left[{-2\sqrt{\lambda} \over 3}(t-t_{a})\right]}
\eqno (3.17)
$$
so that (3.9) becomes $\dot \alpha = \sqrt{\lambda}\phi_{a}^{2}\exp(-
{4\sqrt{\lambda} \over 3}(t-t_{a}))$, which implies
$$
a(t)=a_{I}\exp{\left\{{3\phi_{a}^{2} \over 4}\left[1-
\exp{\left(-{4\sqrt{\lambda} \over 3}
(t-t_{a})\right)}\right]\right\}} .
\eqno (3.18)
$$

Therefore the cosmic scale factor $a(t_{b})$ at the end of this era is
$$
a(t_{b})=a_{I}\exp{\left[{3\phi_{a}^{2} \over 4}\left(1-
{\left({\phi_{b}\over \phi_{a}}\right)}^{2}\right)\right]}
\cong a_{I}\exp{\left({3\phi_{a}^{2} \over 4}\right)}
\eqno (3.19)
$$
if $\phi_{b} \ll \phi_{a}$, where
$$
t_{b}-t_{a}={3 \over 2\sqrt{\lambda}}\log
{\left({\phi_{a} \over \phi_{b}}\right)} .
\eqno (3.20)
$$

Furthermore, if $n=3$ one finds that
$$
a(t)=a_{I}\exp{\left[\phi_{a}^{2}-{\left(\sqrt{\phi_{a}}
-{\sqrt{\lambda} \over 4}(t-t_{a})\right)}^{4}
\right]} .
\eqno (3.21)
$$

At the end of the era during which (3.21) holds one has
$$
t_{b}-t_{a}=4\sqrt{{\phi_{a} \over \lambda}}\left[1-\sqrt{{\phi_{b} \over
\phi_{a}}}\right] .
\eqno (3.22)
$$

Therefore $a(t_{b}) \cong a_{I}\exp(\phi_{a}^{2})$ if $\phi_{b}\ll
\phi_{a}$. When $t_{a}<t\leq t_{b}$, and $t-t_{a}$ is very small, one has for
any $n$ that
$$
a(t)\cong a_{I} \exp(\sqrt{\lambda}\phi_{a}^{n \over
2}(t-t_{a}))\cong {a_{0} \over 2}\exp(\sqrt{\lambda}\phi_{a}^{n \over 2}t)
$$
which is the correct inflationary formula one
would have expected to hold true. It is to be emphasized
that it would be wrong to put
$t_{a}=0$ for simplicity in formulae (3.10)-(3.22).
In fact, $t=0$ is the time at which $\dot \phi$
and $\dot a$ vanish according to (2.9), whereas $t=t_{a}$ is the time at
which $S$ can be approximated by (3.5) up to $t=t_{b}$.
Therefore during the inflationary era $\dot \phi$ and $\dot a$ do not vanish,
but can be computed by means of
(3.10), (3.11), (3.17) and (3.18). This is what also
happens in the massive scalar field model [8],
in which the solutions
corresponding to the oscillatory part of the wavefunction start out with
$\dot \phi(0)=\dot a(0)=0$, and expand exponentially with $\dot \phi =-{m
\over 3}$, $\dot a=ma\mid \phi \mid$.

Finally, when $\phi$ is negative and $n$ is even, (3.5) becomes
$S\cong -{e^{3\alpha}\sqrt{\lambda}{\mid \phi \mid}^{n\over 2} \over 3}$. In
fact, (3.5) is the approximate form of (3.1), which can also
be written in the following way:
$$
S=-{1\over {3\lambda \phi^n}}\Bigr[e^{2\alpha}\lambda\phi^{n} - 1\Bigr]
\sqrt{e^{2\alpha}\lambda \phi^n - 1} .
$$
If we choose the square root on the
right-hand side to be positive, as we already did in the case of positive
values of $\phi$, this can only be obtained introducing $\mid \phi \mid$,
because otherwise $\phi^{n\over 2}$ could be negative when $n$ is even but
${n\over 2}$ is odd. Thus in equation (3.8) $\dot \phi$ is to be replaced by
the time derivative of $\mid \phi \mid$ and in (3.9) $\phi^{n\over 2}$
is to be
replaced by ${\mid \phi \mid}^{n\over 2}$. This implies that, for any
even $n$, when $\phi$ is negative, $\mid \phi \mid$ is a decreasing function
of the time during the inflationary era. Such a behaviour is typical of a
field $\phi$ which starts from a large negative value and finally can reach
zero before starting to oscillate. Thus
the relations (3.10)-(3.22) hold true
provided that one replaces $\phi$ by $\mid \phi \mid$.

So far we have derived solutions to the first-order system
$$
{\partial S
\over \partial a} = {\partial L \over \partial \dot a}
$$
$$
{\partial S \over
\partial \phi} = {\partial L \over \partial \dot \phi}
$$
where $S$ is obtained by analytical continuation
of the Euclidean action for compact 4-metrics and regular matter fields,
and $L$ is the Lagrangian defined in (2.1). In particular,
we have focused our
attention on the case when $S$ takes the approximate form
(3.5). Some important comments are now in order. In our model the full field
equations are the second-order differential
equations (2.2) and (2.3) plus
the Hamiltonian constraint:
$$
a^2 {\dot \phi}^2 - {\dot a}^2 + \lambda a^2 \phi^n - 1 =0 .
\eqno (3.23)
$$

Thus the solution to the full field equations involves three arbitrary
constants. On the other hand, a wave function of the form $Ce^{iS}$ is peaked
about the first integral: $p_{a}={\partial S \over \partial a}$,
$p_{\phi}={\partial S \over \partial \phi}$. This first integral consists
of two first-order ordinary differential
equations (see (3.3) and (3.4))
and so the solution involves just two arbitrary constants. The wave function
is therefore peaked about a set of solutions which are a subset of the general
solution.

Using the Hamilton-Jacobi equation for $S$, we shall show at first that the
solutions of the above-mentioned first integral satisfy the full field
equations exactly.
Finally, we will briefly compare the set of solutions picked out
by the Hartle-Hawking wave function with the general solution.

The Hamilton-Jacobi equation
(2.11) satisfied by $S$ can be cast in the form:
$$
g^{ij}{\partial S \over \partial q^{i}}
{\partial S \over \partial q^{j}}+V=0
\eqno (3.24)
$$
where $g^{ij}={\rm diag}(-1,1)$,
$q^{1}=\alpha$, $q^{2}=\phi$, and V=V($\alpha$,
$\phi$)=$\lambda\phi^{n}e^{6\alpha} - e^{4\alpha}$ is the potential already
defined in section 2. Following Halliwell [21],
let us now differentiate (3.24)
with respect to $q^{k}$. In so doing we get
$$
2g^{ij}{\partial S \over \partial q^{i}}
{\partial ^{2} S \over \partial q^{j} \partial q^{k}}
+{\partial V \over \partial q^{k}}=0 .
\eqno (3.25)
$$

Let us now introduce, again according to [21], the vector
$$
{d\over d\tau}  =g^{ij}{\partial  S \over \partial q^{i}}
{\partial \over \partial q^{j}} .
\eqno (3.26)
$$

The right choice for the parameter $\tau$ is its identification with proper
time: $\tau$ = $\int Ndt$. In fact one can easily check that in so doing one
obtains the well known relation between $\dot \alpha$, $\dot \phi$, and the
momenta $p_{\alpha}$ and $p_{\phi}$. Therefore ${d\over d\tau} =
{d\over dt}$ if $N=1$. The insertion of (3.26) into (3.25)
yields finally
$$
{dp_{k}\over dt}+{1\over 2}{\partial V \over \partial q^{k}}=0
\eqno (3.27)
$$
and (3.27) is just
the form assumed by the field equations. In fact, (2.2) and
(2.3) may also be cast in the first-order form:
$$
\dot p_{\phi}=-{n\lambda \over 2}\phi^{n-1}e^{6\alpha}
\eqno (3.28)
$$
$$
\dot p_{\alpha}=2e^{4\alpha}  -3\lambda\phi^{n} e^{6\alpha}
\eqno (3.29)
$$
where we have used the following formula for the Hamiltonian:
$$
H={1\over 2}(p^{2}_{\phi}-p^{2}_{\alpha})-{1\over 2}e^{4\alpha}
+{\lambda \over 2}\phi^{n}e^{6\alpha} .
\eqno (3.30)
$$
If we now put
$q^{k}$=$\phi$ in (3.27) we obtain (3.28), and if we put $q^{k}$
=$\alpha$ we obtain (3.29).

The task remains now to examine the following point: what is so special about
the solutions picked out by the Hartle-Hawking proposal in comparison to the
general solution? In fact, if this question is not addressed, it is difficult
to see what we have gained by computing the wavefunction for our model.

Indeed, the calculations and the arguments developed so far have shown that
the solutions about which the Hartle-Hawking wavefunction is peaked are
distinguished in that they are inflationary. They have this property in
view of the fact that the scalar field is
initially very large so that $\lambda
\phi^n$ acts like an effective cosmological constant in the early universe,
whereas the initial value of $\dot \phi$ is very small (see (2.9)). However,
a member of the set of general solutions will not always be inflationary.
This is easily understood looking more carefully at (2.2)
and (3.23). Putting as usual $\alpha={\rm ln}(a)$,
these relations imply that
$$
\ddot \alpha= -{\dot \alpha}^2 - 2{\dot \phi}^2 + \lambda\phi^{n} .
\eqno (3.31)
$$

Thus an inflationary solution must be such that
$\dot \alpha={\rm constant}=H>0$,
$\ddot \alpha=0=-H^2 -2{\dot \phi}^2 + \lambda\phi^n$. But a member of the
set of general solutions might well have an initial value of $\dot \phi$
which is very large. Under such a condition, the right-hand side of (3.31)
does not vanish if $(\lambda\phi^n - {\dot \alpha}^2) \ll 2{\dot \phi}^2$.
This implies in turn that no inflation is possible if $\dot \phi$ is initially
so large.

Therefore the Lorentzian Hartle-Hawking trajectories are very peculiar in that
they are singularity free in the past (see (2.8)) and inflationary. The full
comparison of these trajectories with the general solution may be done using
the phase-plane method, generalizing the work
done by Belinski {\it et al} [22].
This analysis may be an interesting problem for further research, but our
analysis already shows the main difference between the two sets of solutions.
\vskip 1cm
\leftline {\bf 4. Minimal conditions for a sufficient inflation}
\vskip 1cm
\noindent
During the inflationary era of the universe, (3.18) becomes
$$
a(t)\cong a_{I}\exp[\sqrt{\lambda}\phi_{a}^{2}(t-t_{a})]\cong
{a_{0} \over 2}\exp(\sqrt{\lambda}\phi_{a}^{2}t) .
\eqno (4.1)
$$

If we require that the inflationary formula (4.1) satisfies the condition
$a(t)\geq a_{0}10^{28}\cong a_{0}\exp{(65)}$ in order to solve the horizon and
flatness problems [23-25],
and if we put $t=\beta t_{b}$, where $\beta
\in ]0,1]$, we find the condition (see (3.20))
$$
\sqrt{\lambda}\phi_{a}^{2}t_{a}+{3 \over 2}\phi_{a}^{2}
\log{\left({\phi_{a} \over
\phi_{b}}\right)}\geq {(65+\log(2))\over \beta}
\eqno (4.2)
$$
in the case of a $\lambda\phi^{4}$ theory. In the same way, we find that the
conditions to be satisfied in the cases $n=3$, $n>4$ are, respectively,
$$
\sqrt{\lambda}\phi_{a}^{3 \over 2}t_{a}+4\phi_{a}^{2}\left(1-\sqrt
{{\phi_{b}\over \phi_{a}}}\right)\geq {(65+\log(2))\over \beta}
\eqno (4.3)
$$
$$
\sqrt{\lambda}\phi_{a}^{n \over 2}t_{a}+{12 \over n(n-4)}
\left({\phi_{a} \over \phi_{b}}\right)^{n \over 2}\phi_{b}^{2}\geq
{(65+\log(2)) \over \beta} .
\eqno (4.4)
$$

However, a thorough inflationary model also has to solve other problems such
as, for example, the origin of the energy density fluctuations. Therefore the
relations (4.2)-(4.4)
are just a part of the minimal requirements to be satisfied by our model.
\vskip 1cm
\leftline {\bf 5. Conclusions}
\vskip 1cm
\noindent
Many authors [20, 22, 24-28],
by using a wide range of techniques, had already
considered the effect produced in the early universe by a term in the scalar
potential of the type $\lambda\phi^{4}$. In this paper we have studied the
general case of a $\lambda\phi^{n}$ theory in a closed FRW minisuperspace
model. We have mainly studied the case when $\phi$ is positive and we have
shown that the case when $\phi$ is negative and $n$ is even can also be taken
into account, provided that one replaces
$\phi$ by $\mid\phi\mid$ in all the
equations of the theory. But if $\phi$ is negative and $n$ is odd,
this does not
give rise to a positive effective cosmological constant in the early universe.

By applying the Hartle-Hawking proposal we have been able to derive
the initial conditions for the solutions to the Lorentzian field equations
(see (2.8) and (2.9)). We have also shown that a suitable choice of
factor ordering in the Wheeler-DeWitt equation (see (2.10)) enables
one to extend to our model a result that is already known in the
case of massive scalar fields: namely, having taken the
Euclidean functional integral over compact 4-metrics and
regular matter fields, the wavefunction is normalizable to one
in the limit of small 3-geometry and it starts to oscillate
when the potential in the Wheeler-DeWitt equation is positive
and $\phi$ is greater than a number of order one. However,
we also emphasized that a more careful calculation and a
deeper understanding of the semiclassical approximation in
quantum cosmology are perhaps needed before taking for granted
that, in general, the wavefunction is a constant on the
boundary of minisuperspace.

Another interesting result is that, with respect to the
massive model and for a given value of $\phi$, there may be an even larger
range of values (i.e. also including smaller values) of the cosmic scale
factor $a(t)$ for which the potential in the Wheeler-De Witt equation is
positive. The initial
conditions (2.8) are indeed in agreement with this fact:
the universe starts off in a non-singular state and the initial value of
$a(t)$ is smaller
than the one for the massive scalar field model, if $\lambda$
is not too small. In the oscillatory region for the wavefunction, we have
taken the phase $S$ of the WKB approximation
to be the analytical continuation
of the Euclidean action for compact 4-metrics and regular matter fields.
The LHH trajectories are thus derived by solving the first-order system
$$
p_{a}={\partial S \over \partial a}={\partial L \over \partial \dot a}
\; \; \; \; \; \; \; \;
p_{\phi}={\partial S \over \partial \phi}={\partial L \over \partial \dot
\phi}
$$
where $L$
is the Lagrangian of the theory (see (2.1)). The solutions of
this system are exact solutions of the field equations; they have vanishing
time derivative in $t=0$ and describe an
exponentially expanding universe at small
values of $(t-t_{a})$, where $t>t_{a}$
and $[0,t_{a}[$ is the time interval during which the field $\phi$
has a constant large value. The Hartle-Hawking wavefunction is thus peaked
about those classical solutions which are inflationary, whereas a member of
the general set of solutions of the full field equations will not always
be inflationary.

The duration of the inflationary era may still be of the order of
$10^{-35}$sec or $10^{-33}$sec (which is very long if compared to the Planck
time) as in the case of the massive
scalar field model, provided that the
parameter $\lambda$ is suitably chosen for any value of $n$. Other important
conditions to be
satisfied are the ones expressed in (4.2)-(4.4) in order to solve
the horizon and flatness problems. Before going over the
matter- or radiation-dominated phase, the universe may expand up to a
maximum value of the order of $\exp({3\phi_{0}^{2} \over n})$ for any $n$ with
respect to the value of $a(t)$ at the beginning of the era when $\exp(-Ht)$
can be neglected in (2.6).
As the time passes, $\phi$ decreases, but the approximate formula
(3.5) from which we have derived most results in section 3
is still approximately
valid as far as $e^{\alpha}$ is very large.

In our opinion it would be particularly interesting to try to
solve numerically the field equations using the initial conditions derived
from the Hartle-Hawking proposal for our model, along the lines of the work
done by Laflamme and Shellard [12]
for the massive model. We think that a
thorough examination of this problem, both in the isotropic and the
anisotropic case, could be of considerable importance in understanding whether
or not the universe described by quantum cosmology will recollapse to a
singularity after a maximum expansion.

Otherwise stated, it may seem quite reasonable that most (or all)
minisuperspace toy models which are able to drive an inflationary era of the
early universe lead to a singularity in the future apart from bouncing
solutions, but so far this is an
unproved conjecture. The proof of this result for $\lambda\phi^{n}$ theories
and other models might show an intriguing link between inflationary models and
the problem of the singularity in the future. In addition, it could be even
more interesting to show that there are some physically meaningful
minisuperspace models for which this result does not hold true.

Finally, we would like to address the attention of the reader to the
following point. If we multiply (2.1) by $-i$ and if
we put $N\rightarrow -iN$,
we obtain a Euclidean action $I_{E}$ which, after the substitutions
$N\rightarrow i$, $a\rightarrow ia$, $\phi \rightarrow i\phi$, becomes
$$
I_{E}= \int \left[{a{\dot a}^{2} \over 2} +
{a \over 2} + {a^{3}{\dot \phi}^{2} \over 2} +
{\lambda a^{3}i^{n}\phi^{n} \over 2}\right]d\tau
$$
which is positive definite for
all values of $n$ of the form $n = 4m$, where $m = 1, 2, 3,...$.
This seems to
suggest that the most interesting $\lambda\phi^{n}$ models are just these
ones, generalizing what was already known in the case of a $\lambda\phi^{4}$
theory [11, 20].
\vskip 1cm
\leftline {\bf Acknowledgments}
\vskip 1cm
\noindent
We are grateful to Peter D'Eath, Marek Demianski, Mauro
Francaviglia, Stephen Hawking, Raymond Laflamme, Jorma
Louko, Giuseppe Marmo and Alexander Zhuk for useful
conversations, and to a referee for other suggestions.
One of us (GE) is indebted to St John's College for
financial support when this work achieved completion.
\vskip 10cm
\leftline {\bf References}
\vskip 1cm
\item {[1]}
Hawking S W 1984 {\it Relativity, Groups and Topology II} ed
B S DeWitt and R Stora (Amsterdam: North-Holland) p 336
\item {[2]}
Hawking S W 1984 {\it Nucl. Phys.} B {\bf 239} 257
\item {[3]}
DeWitt B S 1967 {\it Phys. Rev.} {\bf 160} 1113
\item {[4]}
Hawking S W 1982 {\it Pont. Acad. Sci. Scr. Varia}
{\bf 48} 563
\item {[5]}
Hartle J B and Hawking S W 1983 {\it Phys. Rev.}
D {\bf 28} 2960
\item {[6]}
Hawking S W and Luttrell J C 1984 {\it Phys. Lett.}
{\bf 143B} 83
\item {[7]}
Hawking S W and Luttrell J C 1984 {\it Nucl. Phys.}
B {\bf 247} 250
\item {[8]}
Halliwell J J and Hawking S W 1985 {\it Phys. Rev.}
D {\bf 31} 1777
\item {[9]}
Amsterdamski P 1985 {\it Phys. Rev.} D {\bf 31} 3073
\item {[10]}
Misner C W 1972 {\it Magic Without Magic} ed J Klauder
(San Francisco: Freeman) p 441
\item {[11]}
Hawking S W 1987 {\it 300 Years of Gravitation} ed
S W Hawking and W Israel (Cambridge: Cambridge University
Press) p 631
\item {[12]}
Laflamme R and Shellard P E 1987 {\it Phys. Rev.}
D {\bf 35} 2315
\item {[13]}
Louko J 1987 {\it Phys. Rev.} D {\bf 35} 3760
\item {[14]}
Halliwell J J 1987 {\it Phys. Lett.} {\bf 185B} 341
\item {[15]}
Misner C W, Thorne K S and Wheeler J A 1973
{\it Gravitation} (San Francisco: Freeman) p 507
\item {[16]}
Hawking S W and Wu Z C 1985 {\it Phys. Lett.}
{\bf 151B} 15
\item {[17]}
Schleich K 1985 {\it Phys. Rev.} D {\bf 32} 1889
\item {[18]}
Halliwell J J 1986 {\it Nucl. Phys.} B {\bf 266} 228
\item {[19]}
Shellard E P S 1986 {\it PhD Thesis} University of Cambridge
\item {[20]}
Zhuk A I 1988 {\it Class. Quantum Grav.} to be published
\item {[21]}
Halliwell J J 1987 {\it Phys. Rev.} D {\bf 36} 3626
\item {[22]}
Belinski V A, Grishchuk L P, Khalatnikov I M and
Zel'dovich Ya B 1985 {\it Phys. Lett.} {\bf 155B} 232
\item {[23]}
Guth A H 1981 {\it Phys. Rev.} D {\bf 23} 347
\item {[24]}
Linde A 1984 {\it Rep. Prog. Phys.} {\bf 47} 925
\item {[25]}
Brandenberger R 1985 {\it Rev. Mod. Phys.} {\bf 57} 1
\item {[26]}
Carow U and Watamura S 1985 {\it Phys. Rev.}
D {\bf 32} 1290
\item {[27]}
Guth A H and Pi S Y 1985 {\it Phys. Rev.}
D {\bf 32} 1899
\item {[28]}
Madsen M S 1986 {\it Gen. Rel. Grav.} {\bf 18} 879

\bye